\documentclass[conference,compsoc,english]{IEEEtran}
\usepackage{babel}
\usepackage[inline,shortlabels]{enumitem}
\usepackage{todonotes}
\usepackage{balance}
\ifCLASSOPTIONcompsoc{}
  \usepackage[nocompress]{cite}
\else
  \usepackage{cite}
\fi
\usepackage{siunitx}
\usepackage{graphicx}
\usepackage{array}
\usepackage{url}
\usepackage[hidelinks]{hyperref}

\begin{document}
%
\title{Trying an IP Over NDN Packet Gateway}

\newcommand\copyrighttext{%
  \footnotesize \textcopyright 2021 IEEE. Personal use of this material is permitted.
  Permission from IEEE must be obtained for all other uses, in any current or future 
  media, including reprinting/republishing this material for advertising or promotional 
  purposes, creating new collective works, for resale or redistribution to servers or 
  lists, or reuse of any copyrighted component of this work in other works. 
  DOI: \href{http://dx.doi.org/10.1109/IUCC-CIT-DSCI-SmartCNS55181.2021.00047}{10.1109/IUCC-CIT-DSCI-SmartCNS55181.2021.00047}
}
\newcommand\copyrightnotice{%
\begin{tikzpicture}[remember picture,overlay]
\node[anchor=south,yshift=10pt] at (current page.south) {\fbox{\parbox{\dimexpr\textwidth-\fboxsep-\fboxrule\relax}{\copyrighttext}}};
\end{tikzpicture}%
}

\author{\IEEEauthorblockN{Marielena Márquez-Barreiro,
    Miguel Rodríguez-Pérez
    and Sergio Herrería-Alonso}
  \IEEEauthorblockA{atlanTTic Research Center\\
    Universidade de Vigo\\
    36310 Vigo, Spain\\
    Emails: \url{mmarquez@alumnos.uvigo.es}, \url{miguel@det.uvigo.es} and \url{sha@det.uvigo.es}}} 


\maketitle
\copyrightnotice

\begin{abstract}
  Even though the TCP/IP architecture has served the Internet quite
  satisfactorily during its more than forty years of lifespan, there are doubts
  about whether this host-centric paradigm is well suited for the communication
  patterns of modern applications. These applications mainly need to get
  information pieces without being bothered about where they are located in the
  network. Additionally, the lack of both in-built security mechanisms and
  proper global multicast support may also be a symptom of fundamental problems
  with the classic architecture. Proponents of the novel Information Centric
  Networking (ICN) paradigm assert that the replacement of the TCP/IP
  architecture with one based on the information itself could be the ideal
  solution to all these problems. However, if such a replacement ever takes
  place, it will need the help of some transition mechanisms, for instance, for
  the transmission of legacy IP traffic on top of the new ICN based networks. In
  this paper we design and optimize such an open source IP over ICN transition
  application using the Named Data Networking (NDN) proposal as our target ICN
  network. We performed several tests that confirm that our prototype is able to
  transparently transport IP traffic across a pure NDN network with negligible
  packets losses, low jitter, and good performance.
\end{abstract}



\section{Introduction}%
\label{sec:intro}

Nowadays, digital communications play a key role in almost every area of our
society sustained by the common global telecommunication network known as the
Internet. This network is based on the TCP/IP architecture, a layered protocol
stack that allows applications to set up point-to-point channels to, in essence,
transmit arbitrary information. Though this paradigm has served quite
satisfactorily until now, it does not match well with the new usages and
applications of the Internet, such as digital media, social networking, and
smartphone applications. In fact, the IP requirement of discovering and
specifying the source and the destination of each communication has become a
nuisance for them. Therefore, there is a need to leave the original TCP/IP model
(point-to-point communications, endpoints and addresses) behind and focus on
what is relevant to current users and applications: the content itself.


In the last few years a radically different architecture for the network layer
of the Internet has been gaining increased attention from the research
community. Proposals under the Information Centric Networking (ICN)
umbrella~\cite{kutscher_information-centric_2012} enable consumer-driven
communications where users can directly request information pieces, oblivious to
their location. The \emph{Named Data Networking} (NDN)
project~\cite{zhang_named_2014} is one of the most advanced proposals based on
this architecture. It builds on principles adjusted to the new emerging
communication patterns, thus making the network evolve from the current
host-centric TCP/IP architecture to a more suited data-centric one.
Nevertheless, the fact that all current devices and applications are designed
according to the TCP/IP architecture assures that a potential transition towards
NDN will be really challenging. Certainly, the Internet will have to face a
transition period in which both TCP/IP and NDN models will coexist and,
therefore, some tools will be required to ease the migration process. For
instance, it is expected that the deployment of network gateways that enable the
transport of IP traffic through an NDN
network~\cite{fahrianto_dual-channel_2021,fahrianto_low-cost_2021} will receive
a lot of attention in the next few years.


In this paper we present a new gateway design for efficient IP transmission across NDN
realms. The development and proof of this kind of gateways requires performing
the following set of tasks:
\begin{itemize}
  \item The design of a forwarding protocol that enables the transport of IP packets
        using NDN as the network protocol.
  \item The implementation of an application that captures, processes and
        routes the IP traffic over NDN networks.
  \item The development of a testing environment to validate the gateway
        operations. For example, in this paper we will use a stringent
        streaming application to prove the successful transmission of IP data
        through the gateway.
\end{itemize}
The successful implementation of such an IP over NDN gateway would promote and streamline a
future replacement of the current Internet network layer, thus transforming the present
TCP/IP \emph{communication network} into a true \emph{distribution network}.

The rest of this paper is organized as follows. Section~\ref{sec:related}
provides a description of the main NDN mechanisms and describes some previous
attempts to transmit IP packets through NDN networks. Section~\ref{sec:gateway}
describes the proposed IP over NDN gateway. Experimental results are detailed in
Section~\ref{sec:results}. Finally, Section~\ref{sec:conclusions} summarizes our
conclusions.

\section{Related Work}%
\label{sec:related}

In this Section we will first give a brief description of the NDN architecture. Then, we
will discuss some existing literature about the transport of legacy (IP, TCP)
traffic over NDN networks.

\subsection{Named Data Networking---NDN}%
\label{sec:ndn}

The \emph{Named Data Networking} (NDN) project was founded in 2010 by the
\emph{National Science Foundation} (NSF) with the ambitious goal of becoming the
new universal network layer of the Internet. The main service of the NDN network
layer is fetching data. For this, every piece of data is given a unique
name. This service is quite different from that of the Internet’s original
design, which was delivering packets to a given destination address. This
fundamental change attempts to adjust the network layer to the needs of modern
applications. While the only named entities in the IP packets are the
communication endpoints, NDN removes this restriction so that any distributable
object in an NDN network can be named.

In order to fulfill its service, the NDN network layer employs a communication
protocol with two different type of packets: \emph{Interest} and \emph{Data}.
\emph{Interest} packets are sent by consumer applications to request a piece of named
content from the network. The task of the network layer is to forward these \emph{Interest}
packets to any node capable of returning the requested content. Finally, the
content is returned to the consumers inside a \emph{Data} packet forwarded back by the
NDN routers. Instead of the usual push-model of IP networks, where any node can
enter new traffic in the network, the NDN exhibits a pull-model, as the
consumers are the ones who trigger the communication process, since they are in charge of
creating and sending an \emph{Interest} packet to the network in the first
place.

Despite the fact that NDN is still a developing project, it has already produced
some interesting advances during the past few years. For instance, a formal
protocol specification is available for all users, including a standardization
of the format of NDN packets~\cite{named_data_networking_project_ndn_nodate}, a
detailed description of the NDN link layer~\cite{junxiao_ndnlpv2_nodate} and the
documentation of multiple libraries to encourage the development of new NDN
applications. In addition, the NDN Platform is also available for this purpose,
which includes several tools such as the \emph{NDN Forwarding Daemon}
(NFD)~\cite{noauthor_named_2021}, the \emph{ndn-cxx}
library~\cite{afanasyev_ndn-cxx_nodate} and the
\emph{ndn-tools}~\cite{afanasyev_ndn-tools_nodate}. In fact, there are various
applications that have already been deployed and tested. Among them, there are a
few kinds that must be highlighted: video
streaming~\cite{kulinski_ndnvideo_2012,ghasemi_far_2020}, real-time multiuser
interaction~\cite{gusev_ndn-rtc_2015}, vehicular
networking~\cite{saxena_implementation_2017}, etc. All of them have underlined
and validated the major advantages of NDN over TCP/IP\@:
\begin{enumerate*}[a)]
  \item its design oriented to content distribution;
  \item namespace design based on the particular needs of each application;
  \item robust and data-centric security; and
  \item support for mobility and broadcast \end{enumerate*}.

\subsection{IP Over NDN}%
\label{sec:ipndn}

The research literature about NDN already contains several approaches for the
coexistence and migration of IP networks to NDN ones. According
to~\cite{fahrianto_comparison_2020}, there are several alternative approaches
for such a migration: \begin{enumerate*}[\em a)]
  \item a dual stack approach, akin to the one used for the IPv4 to IPv6
  transition;
  \item a so-called hybrid approach, that performs address translation; and
  finally,
  \item a translation approach that devises gateways able to communicate between
  both stacks. These gateways may function at different levels of the
  communication stack.
\end{enumerate*}
We are interested, in particular, in those translating gateways that
perform network level translation so that any IP traffic can cross an NDN
domain.

A few works have already explored this issue. \cite{luo_ipndn_2018}~proposes
and investigates different approaches to build a gateway. One of their
alternatives is an straightforward network-layer gateway similar to
what we will describe later as our first naive approximation to
the problem in Section~\ref{sec:basic}. However, such an approach suffers from
poor performance and high transmission overhead. Other authors have tried more
specialized gateways. For instance, in~\cite{moiseenko_tcpicn_2016} the authors
describe a proxying gateway for carrying TCP traffic on top of an ICN network.
Another alternative based on translation is presented
in~\cite{fahrianto_dual-channel_2021}. This particular gateway, called
dual-channel translation gateway, provides a privacy-preserving translation
method that uses a name resolution table in order to securely bind an IP packet
with a prefix name. The solution is appropriate for deploying content-centric
applications in the IP realm, but, differently from ours, it provides no mechanisms
for carrying general IP traffic across NDN only cores. Lastly,
in~\cite{fahrianto_low-cost_2021} we find another work that opts for translation
as the approach for IP to NDN migration. In this case, the translation method
utilizes the data payload for bridging the semantic protocol gap in some
possible producer and consumer scenarios. As before, it is assumed that the IP
application is aware of the existence of the translation being carried out.

Taking all this into consideration, we consider that there is still an important
need to fully fill the existing gap in the migration process from IP networks to
NDN ones. And here is where our proposal of an IP over NDN gateway comes into play.

\section{Our IP over NDN Gateway}%
\label{sec:gateway}

\begin{figure}
  \includegraphics[width=\columnwidth]{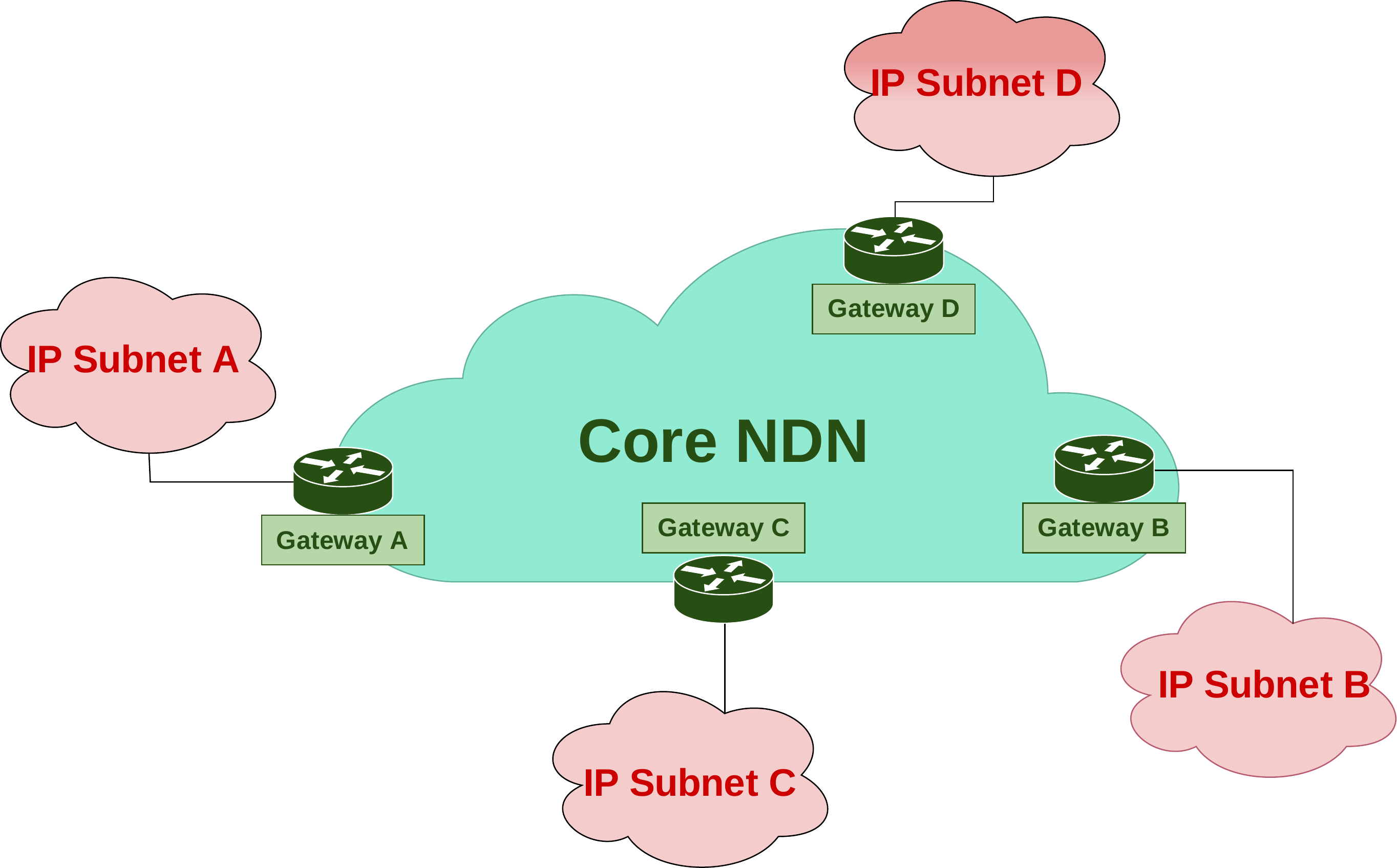}
  \caption{A core NDN network serving as the link layer for four IP subnets.
    Each router located at the edge of the NDN core must run an IP over NDN gateway.}%
  \label{fig:scenario}
\end{figure}

Lets consider an example scenario as the one shown in Fig.~\ref{fig:scenario}
that will serve as a reference for the description of our gateway.\footnote{We
  have published the gateway code as free software, available for download
  at~\url{https://icarus.det.uvigo.es/ndnOverIP/}.} The core of the network is
composed of several routers that only employ the NDN protocol. There are also some
hosts on different IP subnets that want to exchange IP traffic. Each IP
subnet is connected to a gateway router that implements both the NDN and the
IP architectures. In this way, IP packets can travel from one subnet to
another one if the ingress gateway can forward the traffic
across the NDN domain to the egress gateway connected to the destination subnet.
For this scheme to work, each gateway must be the \emph{producer} of a
particular NDN \emph{prefix} related to the address of the IP subnet to which
it belongs. That is, it will produce named \emph{Data} packets for IP datagrams
captured by its IP interface when the destination address resides in an IP
subnet connected to a different gateway. As we are only concerned with the
design of the forwarding protocol, we will assume, for simplicity, that all the
gateways know the routing table which lists the pairs of NDN gateways
identifiers and the addresses of their directly connected IP subnets. This
routing table dictates towards which gateway a specific IP packet must be
forwarded according to its destination address. Thus, all the gateways will know
which IP subnets are reachable through each gateway in the network.

To be able to forward IP traffic across the NDN core, the gateways must
first capture the IP packets coming from the IP subnets they are connected to,
so our tool also includes a traffic capture module implemented with the
\emph{libpcap} library~\cite{tcpdump_group_libpcap_nodate}. This module sets up
each network interface with a capture filter for IP traffic and it creates a
process that waits for new IP packets to arrive. Each captured IP packet  is
processed to obtain its destination address and its length, which will be
important when managing its forwarding through the NDN core. Then, if the
forwarding table contains a next-hop gateway on the NDN core corresponding to
the destination address, the packet is forwarded according to the protocol to be
detailed below.

Once the IP packet arrives as the content of a \emph{Data} packet at the
corresponding egress gateway, our tool injects it
as a raw IP packet in the destination subnet. The packet injection module of
our tool uses RAW \emph{sockets}, as they allow the direct transmission of IP
packets without using any transport protocol and without specifying the link
layer headers. In particular, the RAW sockets configured belong to the
\texttt{AF\_INET} family, which delegates the construction of the link layer
headers to the operating system itself. Furthermore, the \emph{protocol} field
specified during the creation of the sockets is \texttt{IPPROTO\_RAW}, which
means that, whenever the \emph{sendto} function is called, the IP header itself is
contained in its \emph{buffer} parameter. Therefore, our module writes each IP
header manually and the destination address included in it is the one used to
forward the packet.

\subsection{Basic Forwarding Protocol}%
\label{sec:basic}

As it was previously mentioned, the main task of the IP over NDN gateway is to
solve the mismatch between the push model of the IP communication and the pull model
of the NDN architecture. That is, a \emph{Data} packet cannot be transmitted
without a pending \emph{Interest} packet in the network that has previously
requested it. Thus, the design of this protocol must rely on one of the main NDN
principles: communication is always triggered by a consumer.\footnote{The
  question about whether it is possible to send data inside \emph{Interest} packets is a
  recurrent one in ICN forums. It is generally frowned upon as it would blur the
  distinction between both packet types and prevent the correct usage of the
  in-network caches, among other things.} This mechanism is triggered whenever the
gateway captures an IP packet and there is a matching entry for its destination
address in the routing table. As a result, a gateway will not be able to send
data (IP packets in this case) across the NDN core if it has not received a
previous request from another gateway.

\begin{figure}
  \includegraphics[width=\columnwidth]{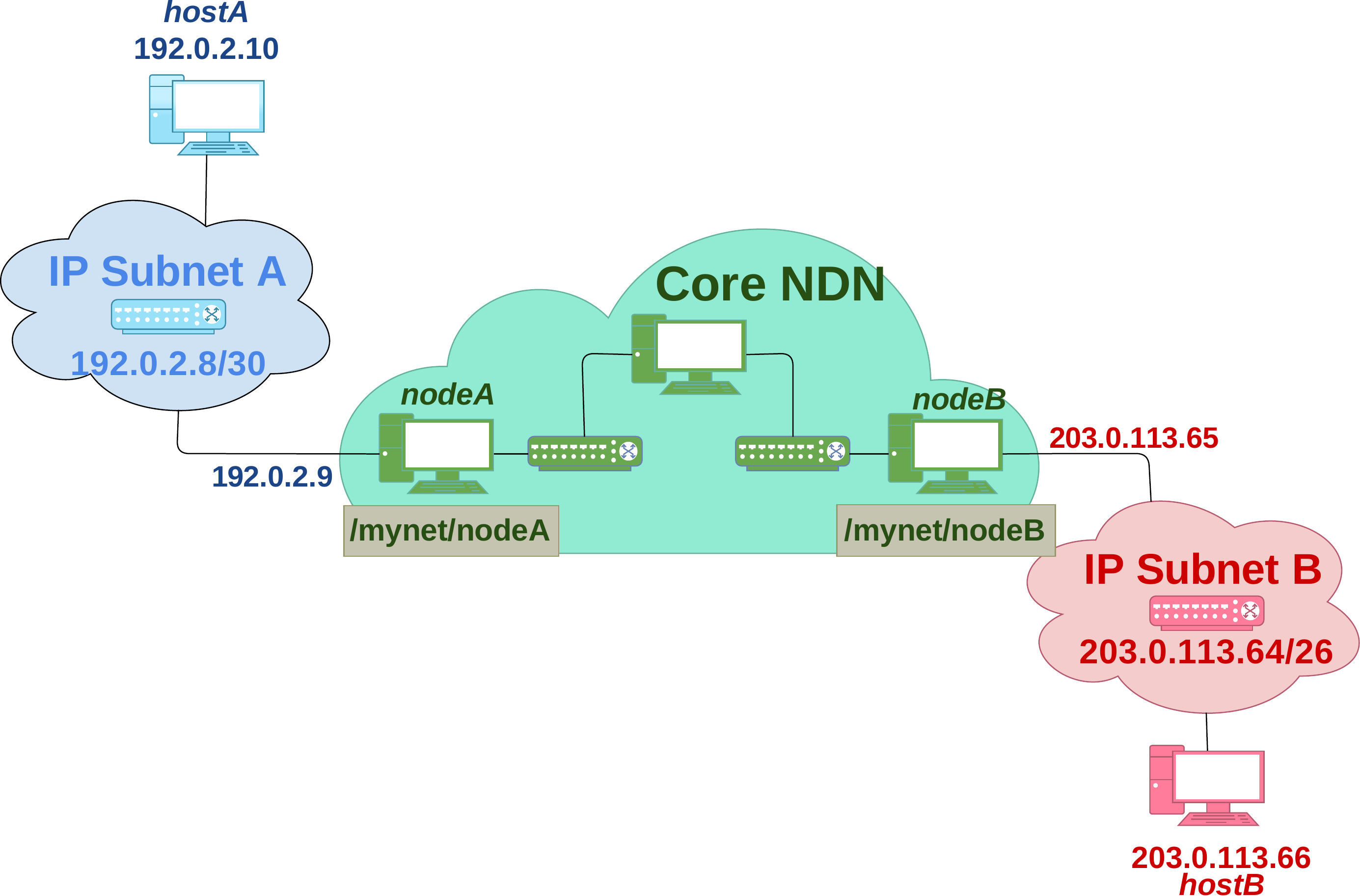}
  \caption{IP over NDN example scenario.}%
  \label{fig:scenario-explanation}
\end{figure}

Having this constraint in mind, we propose a basic IP over NDN protocol
involving three different stages which must be executed for each captured IP
packet. In order to explain this protocol more clearly, we consider the
particular scenario shown in Fig.~\ref{fig:scenario-explanation} in which a host
located in the IP subnet~\emph{B} (referred to as \emph{hostB}, with address
\emph{203.0.113.66}) sends an IP packet to a host located in the IP
subnet~\emph{A} (referred to as \emph{hostA}, with address \emph{192.0.2.10}).
This particular case can be generalized to any scenario in which a host located
in one of the IP subnets sends data to another host located in a different IP
subnet. In the proposed scenario, the process shown in Fig.~\ref{fig:process}
takes place as follows:
\begin{figure}
  \includegraphics[width=\columnwidth]{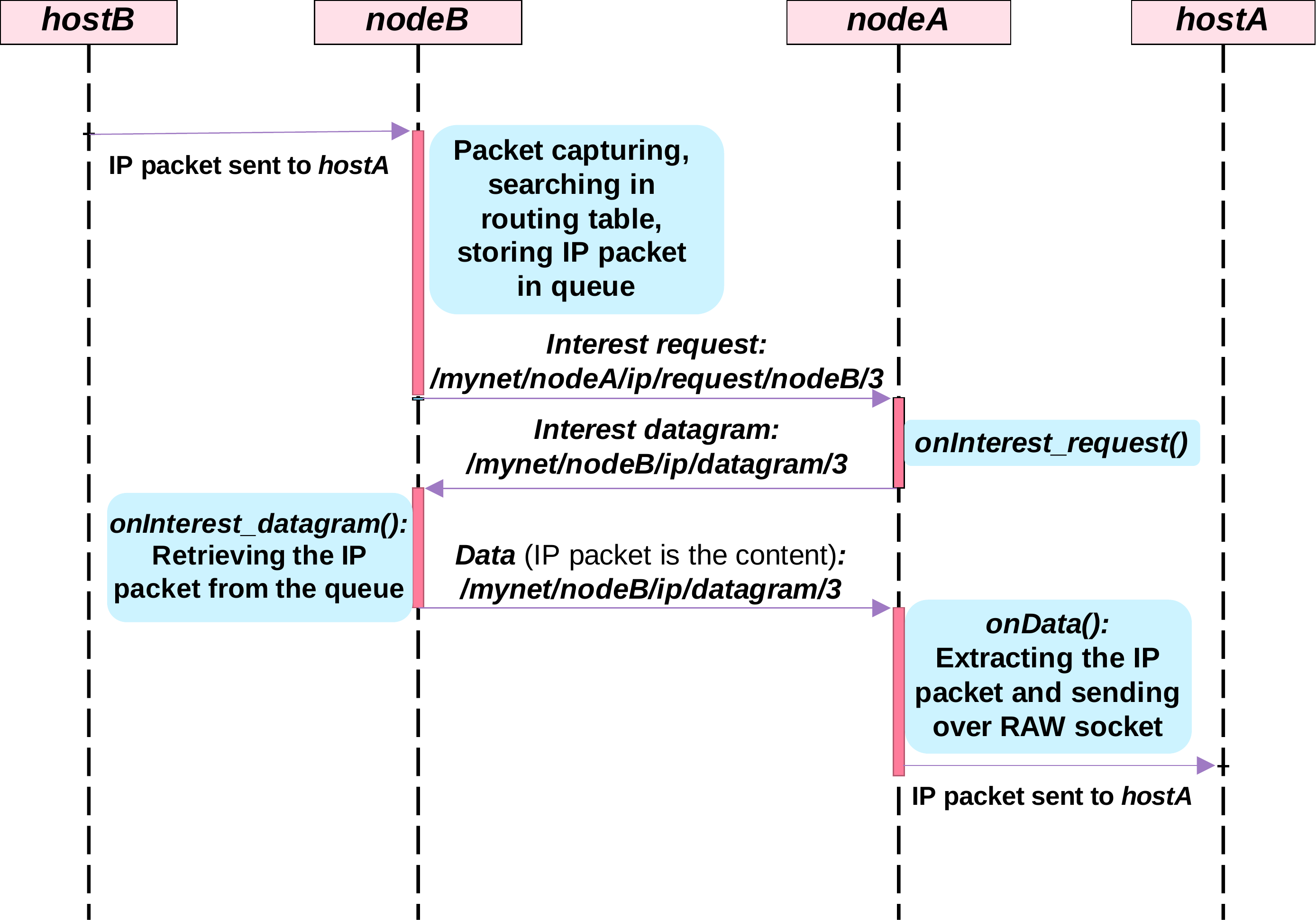}
  \caption{Time diagram of the basic gateway protocol.}%
  \label{fig:process}
\end{figure}
\begin{enumerate}
  \item The \emph{nodeB} gateway captures the IP packet sent by \emph{hostB} and
        searches for an entry in the routing table matching the corresponding IP
        destination address (\emph{192.0.2.10}). Thus, \emph{nodeB} finds out that
        \emph{hostA} is reachable through the \emph{nodeA} gateway. Then,
        \emph{nodeB} assigns a unique sequence number ($seqno$) to the IP packet
        and stores it in its \emph{pending packet queue}. Note that a sequence
        number is required to identify each IP packet within a gateway since this
        could have several IP packets pending to be forwarded.

        The next step is to notify \emph{nodeA} of the existence of a new IP
        packet destined to it. For this, it sends an \emph{Interest} packet
        asking for a specifically crafted name made up of the following strings:
        \begin{enumerate*}[\em a)]
          \item the identifier of the destination gateway obtained from the
          routing table;
          \item the string \texttt{/ip/request/}, where \texttt{ip} identifies
          the gateway process and \texttt{request} specifies the method;
          \item the identifier of the source gateway; and
          \item the sequence number assigned to the IP packet.\end{enumerate*} %
        We call this packet an \emph{Interest Request}. In our example, if
        $seqno=3$, the name in the \emph{Interest Request} would thus render
        \texttt{/mynet/nodeA/ip/request/nodeB/3}.

  \item Thanks to the name in the \emph{Interest Request}, this packet sent
        by \emph{nodeB} is forwarded until it reaches \emph{nodeA}, since there
        is no \emph{Data} stored in the intermediary nodes that satisfies
        it. Then, \emph{nodeA} executes \emph{onInterest\_request()}, a callback
        function registered during its initialization for the NDN prefix
        \texttt{/mynet/$<$gateway identifier$>$/ip/request/}. Within this
        callback, \emph{nodeA} processes the \emph{Interest Request} name and
        extracts the identifier of the source gateway and the sequence number
        assigned to the IP packet. With this information, \emph{nodeA} sends a
        response \emph{Interest} packet, which we will call \emph{Interest Datagram},
        with a name made up as the concatenation of the following substrings:
        \begin{enumerate*}[\em a)]
          \item the identifier of the gateway that sent the \emph{Interest
            Request} and that will receive this \emph{Interest Datagram};
          \item \texttt{/ip/datagram/}, where, again, \texttt{ip} identifies the
          gateway process and \texttt{datagram} specifies the method; and
          \item the sequence number that was extracted from the \emph{Interest
            Request}.
        \end{enumerate*}
        Thus, in our example, the \emph{Interest Datagram} is named as
        \texttt{/mynet/nodeB/ip/datagram/3}.

  \item The \emph{Interest Datagram} is forwarded until it finally reaches
        \emph{nodeB}, since there is no \emph{Data} packet stored in the network
        core that matches its name. Then, \emph{nodeB} executes
        \emph{onInterest\_datagram()}, another callback method also registered
        during the gateway initialization, for the NDN prefix
        \texttt{/mynet/$<$gateway identifier$>$/ip/datagram}. This function
        processes the name of the \emph{Interest Datagram} and extracts the
        sequence number that it carries. As a result, \emph{nodeB} is able to
        identify the IP packet that this \emph{Interest Datagram} authorizes to
        send and retrieves it from the pending packet queue. This IP packet becomes
        the content of the \emph{Data} packet that \emph{nodeB} sends to the
        network as response to the received \emph{Interest Datagram}.

  \item Finally, when \emph{nodeA} receives the \emph{Data} packet sent by
        \emph{nodeB}, it executes the callback \emph{onData()} to extract the IP
        packet that it contains. As a result, \emph{nodeA} processes the IP header
        of this packet in order to know its destination address and length, and
        transmits it to its final destination on the subnet (\emph{hostA} in our
        example).
\end{enumerate}


\subsection{Improved Forwarding Protocol}%
\label{sec:improved}

Although the basic protocol just described in the previous section works
correctly, its performance is not very satisfactory since, as we will show in
Section~\ref{sec:results}, it obtains too low transmission rates and
significant packet losses. Note that, with this protocol, each \emph{Data} packet
enables the transport of only one IP packet. That is, the exchange of NDN
packets among gateways is triggered once by each IP packet that is captured.
Therefore, the transport of each IP packet involves the sending of three NDN
packets through the core, thus greatly limiting the IP transmission rate.

To improve the system performance, the gateways must implement a new version of
the protocol that reduces the number of NDN packets sent across the network
core. For this, we can take advantage of the actual operation of current
applications: IP packets typically belong to a transport layer connection
and thus arrive in bursts or batches. Therefore, the design of the improved
version of the protocol should take this time dependence into account.



Following this approach, the main enhancement in the new version is that now
each \emph{Data} packet that goes through the NDN core may contain \emph{more
  than a single IP datagram} as long as all the datagrams being transported in the
same \emph{Data} packet satisfy that they share the same destination
gateway. Note that, as in the basic protocol, the improved version also requires
the exchange of three NDN packets, but the possibility of forwarding multiple IP
packets into one \emph{Data} packet will permit downsizing the frequency at which
the exchange of the NDN packets occurs. As a result, the gateways must undergo
the following changes:
\begin{itemize}
  \item While in the basic version each gateway manages only one pending packet
        queue, in the new version each gateway must maintain one separate queue
        for each other gateway in the system. Moreover, each of these queues has
        its own independent sequence number space in order to identify its
        corresponding elements unequivocally.
  \item Since the content of each \emph{Data} packet is no longer an individual
        IP packet but a collection of them, the elements of each queue are now
        called \emph{macro-packets}. This term refers to the concatenation of
        multiple IP packets for the same destination gateway. Thus, in the new
        version, the content of each \emph{Data} packet will be one
        \emph{macro-packet}, consisting of a variable number of IP packets
        conditioned by their size. This restriction is set by the NFD itself
        (the canonical implementation of the NDN network layer), since it limits
        the maximum amount of data that can be transported in a \emph{Data}
        packet.
\end{itemize}

These modifications do not require any changes to the capture and processing
module of the incoming IP packets. Also, the format of the exchanged NDN packets
remains unaffected except for the name used in the \emph{Interest
  Datagrams}. Note that, in the new version, this name must also include the
identifier of the source gateway in order to identify the queue where the named
\emph{macro-packet} is stored. The name carried by the \emph{Interest Datagram} in
our example would be thus \texttt{/mynet/nodeB/ip/dagragram/nodeA/3}, for the
macro-packet with sequence number 3 to be sent from the ingress gateway
\emph{nodeB} to the egress gateway \emph{nodeA}.

However, the storage of the captured IP packets becomes more involved, since each gateway
has now to manage multiple queues of \emph{macro-packets}, instead of only one
queue of individual IP packets. The storing process for each captured IP packet
is shown in Fig.~\ref{fig:storing-process}:
\begin{figure}
  \includegraphics[width=\columnwidth]{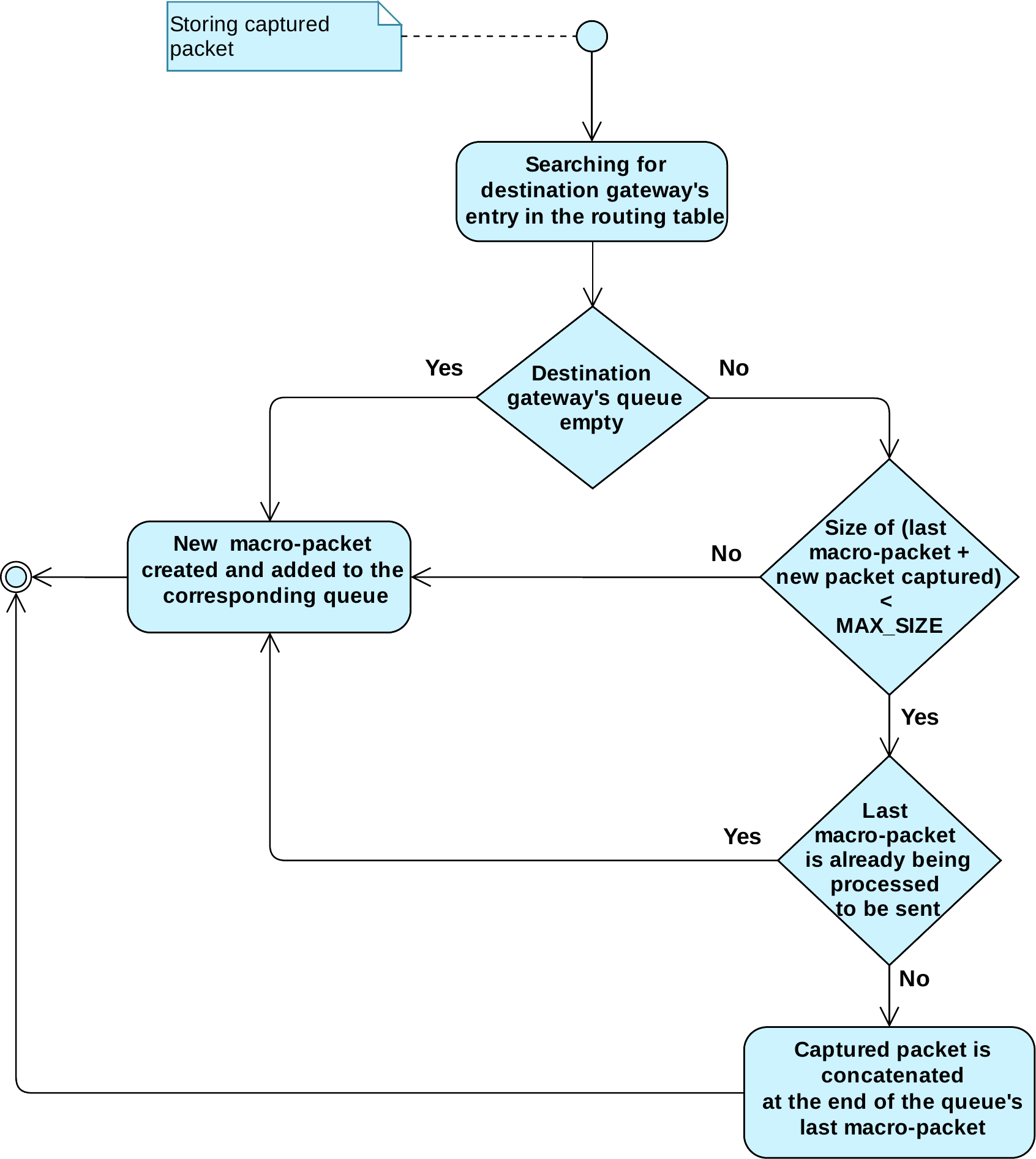}
  \caption{Procedure for storing a newly arrived IP packet at the gateway.}%
  \label{fig:storing-process}
\end{figure}
\begin{enumerate}
  \item Firstly, the gateway finds out the queue in which the IP packet should
        be stored, according to the routing table.
  \item Then, the gateway must determine if a new \emph{macro-packet} has to be
        created in order to store the IP packet just captured. This will only be
        required in three different cases: \begin{enumerate*}[a)]
          \item if the queue is empty;
          \item if the last \emph{macro-packet} in this queue is already being
          processed in order to send it; or
          \item if appending this new IP packet to the end of the last
          \emph{macro-packet} in the queue exceeds the size limit of a \emph{Data}
          packet \end{enumerate*}. Otherwise, it is not required to create a new
        \emph{macro-packet} and the IP packet can be concatenated at the end of
        the last \emph{macro-packet} in the queue.
  \item Finally, the gateway has to decide if it is necessary to send an
        \emph{Interest Request} to the network. The answer to this question will
        be yes only if a new \emph{macro-packet} was created in order to store
        the corresponding IP packet. If this was not the case, we know that an
        \emph{Interest Datagram} will eventually arrive authorizing the sending
        of the \emph{macro-packet} in which this IP packet was stored, since the
        corresponding \emph{Interest Request} was already sent. Thus, there is
        no need to request it again by sending another \emph{Interest Request}.
        Thanks to this logic, we will avoid exchanging three new NDN packets
        through the network core.
\end{enumerate}

All these adjustments in the storing and managing of pending IP packets do not
involve significant changes in how they must be retrieved from the queue. The
only change is that the gateway that receives the \emph{Interest Datagram} needs
not only the sequence number of the \emph{macro-packet} but also the identifier
of the queue where it was stored (information included in the name of the
\emph{Interest Datagram}). The retrieving process itself is analogous to that
taking place in the basic version of the protocol: the \emph{macro-packet}, now
identified by its queue and its sequence number, is retrieved from the queue and
becomes the content of the \emph{Data} packet sent to satisfy the corresponding
\emph{Interest Datagram}.

\begin{figure}
  \includegraphics[width=\columnwidth]{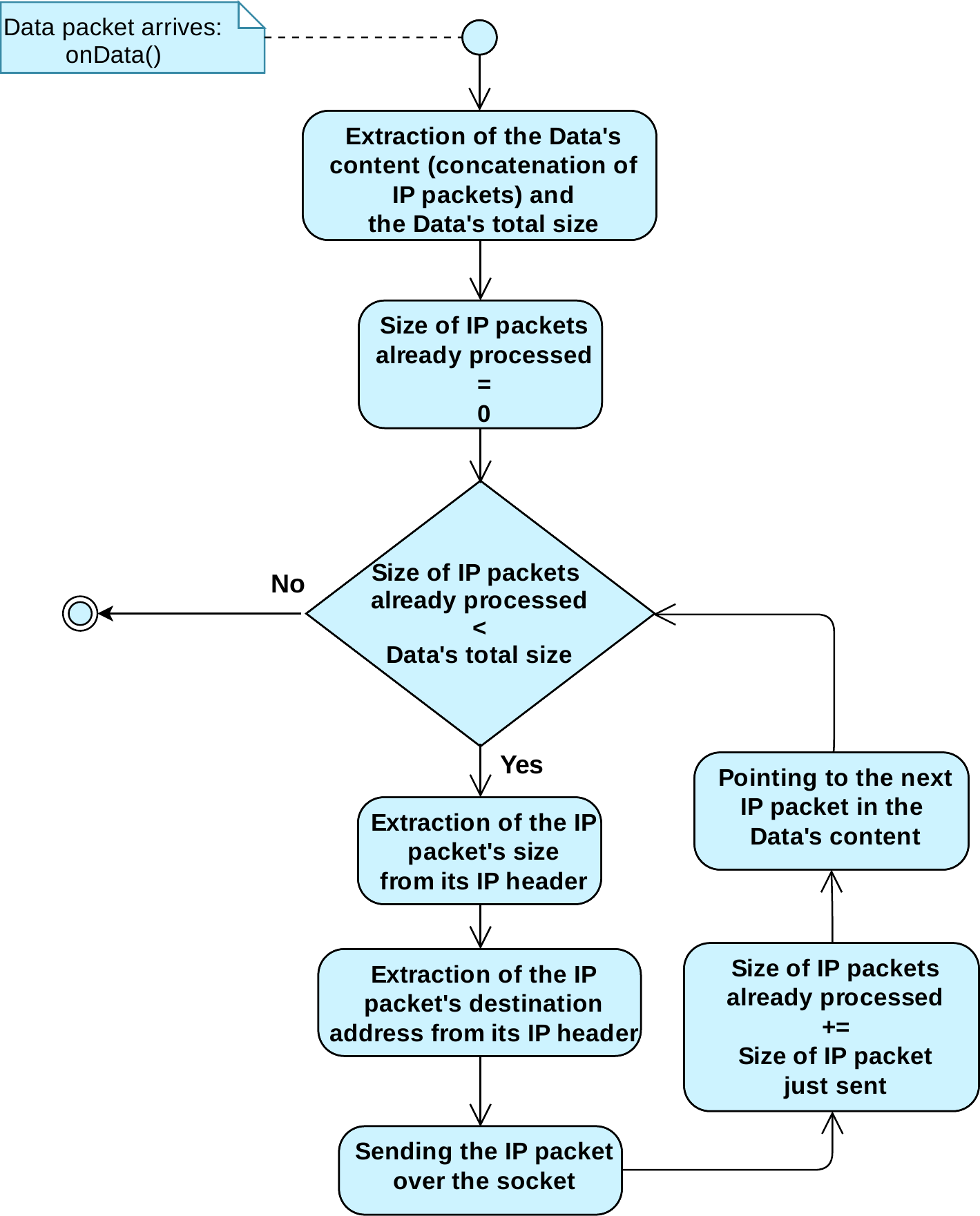}
  \caption{Procedure on arrival of a \emph{Data} packet carrying IP traffic.}%
  \label{fig:particular-operations}
\end{figure}

Finally, note that this improved version also complicates the process executed
by the gateways when they receive a \emph{Data} packet, since its content is no
longer an individual IP packet but rather a set of concatenated datagrams. Thus,
once the \emph{macro-packet} is extracted from the \emph{Data} packet, the
gateway must separate the chained IP packets before sending them individually to
their corresponding IP subnets. This task is accomplished by using the total
size of the content of the \emph{Data} packet and the individual size of each IP
packet (which are obtained from their corresponding headers). The particular
operation done at this point is detailed in
Fig.~\ref{fig:particular-operations}. Thanks to this process, the extracted IP
packets are sent one by one to their IP destination addresses until the end of
the \emph{macro-packet} is reached.

\section{Experimental Results}%
\label{sec:results}

Figure~\ref{fig:scenario-explanation} shows the concrete scenario used to test our
IP over NDN gateway with both the basic and the enhanced protocol versions. As
the diagram illustrates, the system includes two independent IP subnets (subnet
\emph{A} to the left and subnet \emph{B} to the right), both connected to a core
composed of three routers implementing the NDN protocol.
We have performed the same set of tests for both versions of the gateway
protocol to compare their performance. We used the NFD daemon in the NDN
routers. The official C++ NDN library
(\emph{ndn-cxx}~\cite{afanasyev_ndn-cxx_nodate}) was used to implement the main
code of the gateway. For the link level connection between the three NDN routers
we explored three different technologies:
\begin{enumerate*}[\em a)]
  \item using Ethernet directly as the NDN face,\footnote{Generalization of the
    concept of network interface used in the NDN architecture. The term
    \emph{face} refers to an abstraction of any communication channel that the
    NFD may use for forwarding NDN packets, since the exchange of packets is
    done not only over physical network interfaces but also directly between
    applications located in the same machine.} operating both multicast and
  unicast faces with their corresponding MAC addresses;
  \item using the IPv6 support; and
  \item using an IPv4 configuration either with \emph{udp4 multicast} interfaces
  or point-to-point interfaces \end{enumerate*}. After configuring and
implementing all these alternatives, we finally chose the IPv4 configuration
with routes configured over UDP \emph{multicast} faces due to its simplicity.
In any case, note that, for our tests, the gateway performance does not depend
on the link layer technology that much. The configuration of the routes over
these faces was statically done using NFD configuration commands.

Initially, we performed some basic tests using the \emph{ping} command, the
\emph{Wireshark} packet analyzer and the \emph{dissect-wireshark} tool (included
in the \emph{ndn-tools}) to verify that the capture, the transport and the
delivery of IP packets between \emph{hostA} and \emph{hostB} were done properly.

Then, to gain confidence on the gateway implementation and to confirm that it
was also capable of dealing with bursts and the transport of other protocols
besides ICMP (the only one used by the \emph{ping} command), we performed some
additional experiments with \emph{iperf}, a tool that can create data streams
between two endpoints in one or both directions.\footnote{The tests with
  \emph{iperf} were performed executing the server in \emph{hostA} and the client
  in \emph{hostB} and vice versa.} With these experiments, we demonstrated that
our gateway operates successfully even when the hosts transmit large amounts of
data using different transport protocols such as TCP or UDP\@.

\begin{figure}
  \includegraphics[width=\columnwidth]{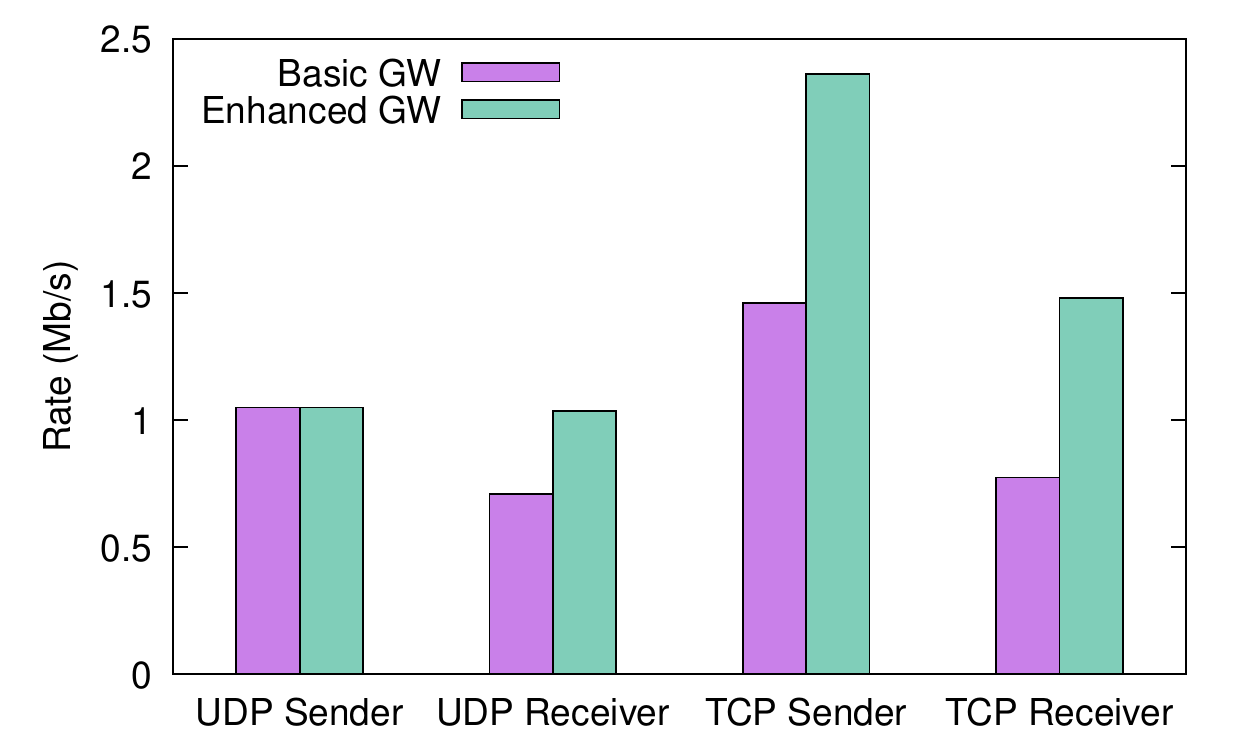}
  \caption{Transmission rates obtained in the \emph{iperf} experiments.}%
  \label{fig:performance-results}
\end{figure}

Additionally, we also used the \emph{iperf} tests to analyze the performance
provided by both the proposed forwarding protocols (the basic protocol and the
enhanced one). Figure~\ref{fig:performance-results} shows the transmission rates
obtained with UDP and TCP when using the basic gateway protocol or the improved
one.\footnote{Each \emph{iperf} experiment was repeated 10 times to obtain
  reliable measurements. The mean and the 95\% confidence interval (CI) of the
  transmission rate at both the sender and receiver sides were calculated, but CIs
  are not shown in the graph since all of them are small and just clutter the
  figure.} As expected, the rate achieved with the improved version of the gateway
is significantly higher. Note that the \emph{iperf} UDP sender is unaware of packet
losses and thus obtains the same throughput regardless of what happens to
the transmitted traffic.

We also measured other important performance parameters, such as the jitter and
the packet loss rate, when sending UDP traffic. Thus, we observed that the
average jitter when using the improved version of the gateway is reduced from
\SIrange{13.8}{8.6}{\ms}. Moreover, with the improved protocol, no packet losses
were experienced, compared to the \SI{7.5}{\percent} packet loss rate measured
with the basic one.

Finally, we performed a visual demonstration with video streaming. This is a
good test to check the proper operation of our gateway, as both jitter and packet losses
manifest clearly in the resulting reproduction. In particular, we ran a
streaming application between \emph{hostA} and \emph{hostB}. In order to perform
this test, \emph{ffmpeg} was used to serve the simultaneous streaming videos
while VLC clients were used to display them.\footnote{Screen captures from the
  test can be visualized at the project web site at
  \url{https://icarus.det.uvigo.es/ndnOverIP/}.} Two specific tests were performed
for each version of the gateway protocol:
\begin{enumerate*}[\em a)]
  \item \emph{hostA} serves a video while \emph{hostB} displays it through a
  VLC client. Simultaneously, \emph{hostB} serves a different video, which is
  displayed by \emph{hostA} through a VLC client as
  well;
  \item \emph{hostB} simultaneously serves two different videos and \emph{hostA}
  displays both of them using two VLC
  clients.\end{enumerate*} %
Both tests were recorded and, as expected, the results were consistent with
those obtained using \emph{iperf}. It can be clearly seen that, when the basic
version of the gateway protocol is used, the video playback is choppy,
stuttering continuously. In contrast, with the improved version, the video
playback is smooth and experiences no pauses nor image corruption artifacts.
Therefore, these tests show that multimedia data can be successfully transmitted
through the gateways, with a good user experience when the improved protocol is
used.

\section{Conclusions}%
\label{sec:conclusions}

The main intention of this work was to contribute to adjust the current network
layer to today's dominant use of the Internet as a content distribution network
rather than as a communication one. For this, we have presented in this paper a
novel design for IP over NDN gateways that includes a new forwarding protocol that enables
the efficient and transparent transport of IP packets through an NDN network.
The proposed gateway design has been implemented and evaluated in a real-world
scenario using the latest available version of the NFD software and standard
tools like \emph{iperf} as well as multimedia traffic with the help of
\emph{VLC}. Performance results revealed that the developed tool is completely
functional and that it can be successfully used to migrate from IP networks to
NDN ones. Moreover, the analysis of these results, both quantitative
(\emph{iperf} results) and qualitative (streaming application results), shows
that our IP over NDN gateway offers a good user experience thanks to the new
efficient forwarding protocol proposed.

As future work, we will explore the following research areas:
\begin{enumerate}
  \item We plan to use \emph{ChronoSync}~\cite{zhu_lets_2013} when configuring
        the routing tables at the gateways. \emph{ChronoSync} is a protocol
        employed to synchronize the state of a dataset among a distributed group
        of NDN users. Therefore, it would allow all the gateways in the network
        to automatically synchronize their routing tables instead of loading it
        individually from a local configuration file during their
        initialization.
  \item We will also explore the possibility that our gateways use \emph{multicast}
        traffic support, taking advantage of the in-network storage that the NDN
        architecture inherently offers. Note that NDN routers can store \emph{Data}
        packets in their \emph{Content Store} temporarily, using it as a cache memory.
\end{enumerate}



\ifCLASSOPTIONcompsoc{}
\section*{Acknowledgments}
\else
\section*{Acknowledgment}
\fi

This publication is part of the research project PID2020-113240RB-I00, financed by MCIN/ AEI/10.13039/501100011033.

\bibliographystyle{IEEEtran}

\bibliography{IEEEabrv,ipndn}

\begin{thebibliography}{10}
\providecommand{\url}[1]{#1}
\csname url@samestyle\endcsname
\providecommand{\newblock}{\relax}
\providecommand{\bibinfo}[2]{#2}
\providecommand{\BIBentrySTDinterwordspacing}{\spaceskip=0pt\relax}
\providecommand{\BIBentryALTinterwordstretchfactor}{4}
\providecommand{\BIBentryALTinterwordspacing}{\spaceskip=\fontdimen2\font plus
\BIBentryALTinterwordstretchfactor\fontdimen3\font minus
  \fontdimen4\font\relax}
\providecommand{\BIBforeignlanguage}[2]{{%
\expandafter\ifx\csname l@#1\endcsname\relax
\typeout{** WARNING: IEEEtran.bst: No hyphenation pattern has been}%
\typeout{** loaded for the language `#1'. Using the pattern for}%
\typeout{** the default language instead.}%
\else
\language=\csname l@#1\endcsname
\fi
#2}}
\providecommand{\BIBdecl}{\relax}
\BIBdecl

\bibitem{kutscher_information-centric_2012}
\BIBentryALTinterwordspacing
D.~Kutscher, B.~Ohlman, and D.~Oran, ``Information-{Centric} {Networking}
  {Research} {Group},'' Apr. 2012, publisher: Internet Research Task Force.
  [Online]. Available: \url{https://irtf.org/icnrg}
\BIBentrySTDinterwordspacing

\bibitem{zhang_named_2014}
L.~Zhang, A.~Afanasyev, J.~Burke, V.~Jacobson, K.~Claffy, P.~Crowley,
  C.~Papadopoulos, L.~Wang, and B.~Zhang, ``Named data networking,''
  \emph{Computer Communication Review}, vol.~44, no.~3, pp. 66--73, 2014.

\bibitem{fahrianto_dual-channel_2021}
F.~Fahrianto and N.~Kamiyama, ``The {Dual}-{Channel} {IP}-to-{NDN}
  {Translation} {Gateway},'' in \emph{2021 {IEEE} {International} {Symposium}
  on {Local} and {Metropolitan} {Area} {Networks} ({LANMAN})}, Jul. 2021, pp.
  1--2, iSSN: 1944-0375.

\bibitem{fahrianto_low-cost_2021}
------, ``A {Low}-{Cost} {IP}-to-{NDN} {Translation} {Gateway},'' in \emph{2021
  {IEEE} 22nd {International} {Conference} on {High} {Performance} {Switching}
  and {Routing} ({HPSR})}, Jun. 2021, pp. 1--5, iSSN: 2325-5609.

\bibitem{named_data_networking_project_ndn_nodate}
\BIBentryALTinterwordspacing
{Named Data Networking Project}, ``{NDN} packet format specification,'' version
  0.3. [Online]. Available:
  \url{https://named-data.net/doc/NDN-packet-spec/current/intro.html}
\BIBentrySTDinterwordspacing

\bibitem{junxiao_ndnlpv2_nodate}
\BIBentryALTinterwordspacing
S.~Junxiao, A.~Afanasyev, D.~Pesavento, D.~Newberry, K.~Schneider, and
  T.~Liang, ``{NDNLPv2} --- {NFD}.'' [Online]. Available:
  \url{https://redmine.named-data.net/projects/nfd/wiki/NDNLPv2}
\BIBentrySTDinterwordspacing

\bibitem{noauthor_named_2021}
\BIBentryALTinterwordspacing
``Named {Data} {Networking} {Forwarding} {Daemon},'' Apr. 2021, original-date:
  2014-01-27T07:43:57Z. [Online]. Available:
  \url{https://github.com/named-data/NFD}
\BIBentrySTDinterwordspacing

\bibitem{afanasyev_ndn-cxx_nodate}
\BIBentryALTinterwordspacing
A.~Afanasyev, Y.~Yu, J.~Shi, J.~Thompson, and Z.~Zhu, ``ndn-cxx: {NDN} {C}++
  library with {eXperimental} {eXtensions}.'' [Online]. Available:
  \url{https://named-data.net/doc/ndn-cxx/current/}
\BIBentrySTDinterwordspacing

\bibitem{afanasyev_ndn-tools_nodate}
\BIBentryALTinterwordspacing
A.~Afanasyev and P.~Davide, ``ndn-tools: {NDN} {Essential} {Tools}.'' [Online].
  Available: \url{https://github.com/named-data/ndn-tools}
\BIBentrySTDinterwordspacing

\bibitem{kulinski_ndnvideo_2012}
D.~Kulinski and J.~Burke, ``\BIBforeignlanguage{en}{{NDNVideo}: {Random}-access
  {Live} and {Pre}-recorded {Streaming} using {NDN}},'' {UCLA}, Technical
  {Report} NDN-0007, Sep. 2012.

\bibitem{ghasemi_far_2020}
\BIBentryALTinterwordspacing
C.~Ghasemi, H.~Yousefi, and B.~Zhang, ``Far {Cry}: {Will} {CDNs} {Hear} {NDN}'s
  {Call}?'' in \emph{Proceedings of the 7th {ACM} {Conference} on
  {Information}-{Centric} {Networking}}, ser. {ICN} '20.\hskip 1em plus 0.5em
  minus 0.4em\relax New York, NY, USA: Association for Computing Machinery,
  Sep. 2020, pp. 89--98. [Online]. Available:
  \url{https://doi.org/10.1145/3405656.3418708}
\BIBentrySTDinterwordspacing

\bibitem{gusev_ndn-rtc_2015}
\BIBentryALTinterwordspacing
P.~Gusev and J.~Burke, ``{NDN}-{RTC}: {Real}-{Time} {Videoconferencing} over
  {Named} {Data} {Networking},'' in \emph{Proceedings of the 2nd {ACM}
  {Conference} on {Information}-{Centric} {Networking}}, ser. {ACM}-{ICN}
  '15.\hskip 1em plus 0.5em minus 0.4em\relax New York, NY, USA: Association
  for Computing Machinery, Sep. 2015, pp. 117--126. [Online]. Available:
  \url{https://doi.org/10.1145/2810156.2810176}
\BIBentrySTDinterwordspacing

\bibitem{saxena_implementation_2017}
\BIBentryALTinterwordspacing
D.~Saxena, V.~Raychoudhury, and C.~Becker, ``Implementation and {Performance}
  {Evaluation} of {Name}-based {Forwarding} {Schemes} in {V}-{NDN},'' in
  \emph{Proceedings of the 18th {International} {Conference} on {Distributed}
  {Computing} and {Networking}}, ser. {ICDCN} '17.\hskip 1em plus 0.5em minus
  0.4em\relax New York, NY, USA: Association for Computing Machinery, Jan.
  2017, pp. 1--4. [Online]. Available:
  \url{https://doi.org/10.1145/3007748.3007766}
\BIBentrySTDinterwordspacing

\bibitem{fahrianto_comparison_2020}
F.~Fahrianto and N.~Kamiyama, ``Comparison of {Migration} {Approaches} of
  {ICN}/{NDN} on {IP} {Networks},'' in \emph{2020 {Fifth} {International}
  {Conference} on {Informatics} and {Computing} ({ICIC})}, Nov. 2020, pp. 1--7.

\bibitem{luo_ipndn_2018}
S.~Luo, S.~Zhong, and K.~Lei, ``{IP}/{NDN}: {A} multi-level translation and
  migration mechanism,'' in \emph{{NOMS} 2018 - 2018 {IEEE}/{IFIP} {Network}
  {Operations} and {Management} {Symposium}}, Apr. 2018, pp. 1--5, iSSN:
  2374-9709.

\bibitem{moiseenko_tcpicn_2016}
\BIBentryALTinterwordspacing
I.~Moiseenko and D.~Oran, ``{TCP}/{ICN}: {Carrying} {TCP} over {Content}
  {Centric} and {Named} {Data} {Networks},'' in \emph{Proceedings of the 3rd
  {ACM} {Conference} on {Information}-{Centric} {Networking}}, ser. {ACM}-{ICN}
  '16.\hskip 1em plus 0.5em minus 0.4em\relax New York, NY, USA: Association
  for Computing Machinery, Sep. 2016, pp. 112--121. [Online]. Available:
  \url{https://doi.org/10.1145/2984356.2984357}
\BIBentrySTDinterwordspacing

\bibitem{tcpdump_group_libpcap_nodate}
\BIBentryALTinterwordspacing
{Tcpdump Group}, ``Libpcap.'' [Online]. Available:
  \url{https://www.tcpdump.org/index.html#documentation}
\BIBentrySTDinterwordspacing

\bibitem{zhu_lets_2013}
Z.~Zhu and A.~Afanasyev, ``Let's {ChronoSync}: {Decentralized} dataset state
  synchronization in {Named} {Data} {Networking},'' in \emph{2013 21st {IEEE}
  {International} {Conference} on {Network} {Protocols} ({ICNP})}, Oct. 2013,
  pp. 1--10, iSSN: 1092-1648.

\end{thebibliography}
\balance{}

\end{document}